\documentclass[11pt,onecolumn,superscriptaddress,showpacs,aps,pre]{revtex4-2}
\usepackage[english]{babel}
\usepackage[dvips]{graphics}
\usepackage{graphicx,epsfig}
\usepackage{amsmath}
\usepackage{amsfonts}
\usepackage{amssymb}
\usepackage{xcolor}
\usepackage{multirow}
\usepackage[normalem]{ulem}
\usepackage{booktabs}
\usepackage{float}
\usepackage{makecell}
\usepackage{array}
\usepackage{url}
\usepackage[breaklinks]{hyperref}
\usepackage{placeins}  
\usepackage{amsthm} 
\usepackage{tabularx} 
\usepackage{bm}
\usepackage{lineno}
\usepackage{animate} 
\usepackage{multimedia} 
\usepackage{hyperref}
\hypersetup{
    breaklinks=true,
    colorlinks=true,
    linkcolor=blue,
    urlcolor=blue
}

\newcommand{\CB}[1]{\textcolor{blue}{#1}}
\definecolor{medpurple}{RGB}{128,0,128}

\newcommand{\FinSTOD}{FinSTOD}

\newcommand{\TA}{$\mathcal{T}_A$}
\newcommand{\TB}{$\mathcal{T}_B$}

\newcommand{\C}{\mathcal{C}}

\theoremstyle{definition} 
\newtheorem{definition}{Definition}

\begin{document}

\title{Strong Trajectorial Ontological Differentiation:
A novel
approach to unravel phase-space structures}

\author{P. García-Cuadrillero}
\email{pablo.garcia.cuadrillero@upm.es}
\affiliation{Grupo de Sistemas Complejos, Departamento de Matemática Aplicada, Universidad Politécnica de Madrid, Avenida Juan de Herrera 6, 28040 Madrid, España.}

\author{J. A. Capitán}
\email{ja.capitan@upm.es}
\affiliation{Grupo de Sistemas Complejos, Departamento de Matemática Aplicada, Universidad Politécnica de Madrid, Avenida Juan de Herrera 6, 28040 Madrid, España.}

\author{F. Revuelta}
\email{fabio.revuelta@upm.es (corresponding author)}
\affiliation{Grupo de Sistemas Complejos, Escuela Técnica Superior de Ingeniería Agronómica, Alimentaria y de Biosistemas, Universidad Politécnica de Madrid, Avenida Puerta de Hierro 2-4, 28040 Madrid, España.}


\begin{abstract}
The identification of invariant objects and Lagrangian coherent structures is
a cornerstone of dynamical systems.
As a consequence,
several diagnostic indicators have been established over time, 
such as
the fast Lyapunov indicator, 
the finite-time Lyapunov exponent, 
and Lagrangian descriptors, among others.
In this work, we introduce the
Strong Trajectorial Ontological Differentiation (STOD)
as a novel 
tool to identify phase-space structures.
Unlike other indicators,
STOD does not rely on the study of the tangent flow;
instead, it identifies phase-space structures by comparing trajectories through a component-wise cancellation process inspired on 
the Ontological Differentiation (OD)
that was originally developed for lexical networks
[\href{https://doi.org/10.1103/148p-69wg}{P. Garc\'ia-Cuadrillero, F. Revuelta, J. A. Capitán, Phys. Rev. E \textbf{113}, 014305 (2026)}].
By applying a reversed-time version of STOD ({\FinSTOD})
to five paradigmatic autonomous and
non-autonomous systems of increasing complexity,
we show the excellent performance of this indicator
in the identification of phase-space structures,
adding a new useful tool to the chaotic toolbox.

\end{abstract}

\maketitle

\section{Introduction}

Chaotic motion is ubiquitous in Nature~\cite{Strogatz24}.
It is present in diverse systems, ranging from
the tiny scales of atoms and molecules \cite{Reichl04}
to the vast expanses of celestial mechanics \cite{Celletti10}.
In short,
chaos is created by the
nonlinearities in the equations of motion (EoM), which induce a high sensitivity to
initial conditions (ICs),
leading to divergent dynamical behaviors~\cite{LL10}.
However, the presence of nonlinear EoM is a necessary but not sufficient condition for the onset of chaos. Our understanding of this distinction has matured significantly through advances in dynamical systems theory~\cite{Arnold06, Wiggins2003}, which have provided a geometric framework for describing nonlinear evolution. In this context, the topology of the phase space dictates the qualitative behavior of the system.

In autonomous systems, regular motion is constrained to the surfaces of invariant tori. According to the Kolmogorov-Arnold-Moser (KAM) theorem~\cite{Arnold06},
these tori may undergo structural breakdown as a perturbation parameter is increased,
providing a formal route to chaotic motion.
Chaotic motion is not fully \emph{erratic}:
it is governed by invariant manifolds that, under specific conditions, act as dynamical separatrices. Furthermore, periodic orbits (POs) often persist as the backbone of the chaotic domain. As dictated by the Poincaré–Birkhoff theorem~\cite{LL10, Strogatz24, Ottino1989}, bifurcations can cause these POs to emerge, vanish, or shift in stability, significantly increasing the topological complexity of the system.

The systematic distinction between regular and irregular motion in systems with two degrees of freedom (2-DoF)
originated with the pioneering work of Henri Poincaré, who introduced the celebrated Poincaré surface of section during his analysis of the restricted three-body problem~\cite{Poincare1892}.
In periodically-driven systems, the Poincaré map can still be recovered by sampling the dynamics at the frequency of the external modulation.
Crucially,
normally hyperbolic invariant manifolds (NHIMs)
and their associated stable and unstable manifolds
persist under small perturbations,
remaining as the phase-space cornerstones~\cite{Fenichel71}.

Conversely,
the direct application of 
the Poincaré surface of section to higher-dimensional or aperiodic systems is often limited 
as
the \emph{rules} of motion shift constantly, rendering static geometric sections insufficient \cite{Haller2001}. In these regimes,
classical invariant structures may become analytically intractable or cease to exist in a traditional sense
due to their inherent asymptotic time behavior.
To address this, Lagrangian Coherent Structures (LCS) have emerged as a solid framework to characterize transport and mixing by partitioning the phase space into dynamically distinct regions over finite-time windows~\cite{Haller2015, Peacock2013}.

By acting as the time-dependent analogues of invariant manifolds, LCS have become central to modeling diverse natural phenomena, from pollutant dispersal in the ocean \cite{Prants2025} and larval connectivity in marine ecosystems \cite{SerGiacomi2021} to the transport of atmospheric tracers \cite{Boffetta2001}. Ultimately, identifying these structures reveals the \emph{hidden skeleton} that governs the underlying transport dynamics
, since they act as effective transport barriers
with negligible flux through them~\cite{Shadden2005}.
Consequently, the pursuit of novel techniques to resolve these phase-space structures remains a fundamental challenge, continuously stimulating new generations of researchers.
%
That is the case of the finite-time indicators 
developed over the last decades,
which can unravel
phase-space structures using relatively short integration times.
Among the most popular tools currently employed in the literature to bridge this gap are the finite-time Lyapunov exponent (FTLE)~\cite{Haller2001, Boffetta2001, Shadden2005} and the finite-time size Lyapunov exponent (FSLE)~\cite{Aurell1997}.
These are particularly robust for identifying Lagrangian Coherent Structures (LCS) and the boundaries of fluid mixing. In a similar vein, the fast Lyapunov indicator (FLI)~\cite{Froeschle1997, Guzzo2023} was designed to distinguish between resonant, ordered, and chaotic motion with high computational efficiency.
Other indices that also focus on the evolution of tangent space, such as the smaller alignment index (SALI)~\cite{Skokos01, Benitez15}
and its successor, the generalized alignment index (GALI)~\cite{Skokos2007}, which utilize the alignment of deviation vectors to rapidly detect the dimensionality of invariant tori. Furthermore, the mean exponential growth factor of nearby orbits (MEGNO)~\cite{Cincotta03, Maffione2011} provides a time-averaged perspective on stability, making it a standard choice for long-term simulations in celestial mechanics.

Beyond the mentioned indicators,
other mathematically rigorous frameworks have emerged to provide a deeper characterization of transport. The variational theory of LCS \cite{Haller2011, Farazmand12} established a robust geometric foundation by defining these structures as stationary surfaces of the strain action, ensuring their objectivity under observer transformations. Complementing this, mesochronic plots~\cite{Mezic2010} utilize an ergodic-theoretic approach—specifically through the spectral properties of the Koopman operator-to partition the phase space based on time-averaged dynamical signatures. Finally, transfer operator-based methods~\cite{Froyland2009} offer a probabilistic perspective, identifying \emph{coherent sets} as regions of the phase space that remain minimally dispersive over a finite-time interval.

While the aforementioned methodologies have transitioned the field from simple chaos detection to the comprehensive mapping of the \emph{skeletons} governing transport and mixing, they often require significant mathematical overhead. Most standard indicators, such as FTLE, FLI, or SALI,
rely on the evolution of the tangent space, where the stretching or separation of infinitesimally close trajectories is measured
(usually by linearizing the dynamics)
over a finite-time interval. In contrast, Lagrangian Descriptors (LD) \cite{Madrid09, Mancho2013, Lopesino17, Revuelta19, Revuelta21, Revuelta23, Abdallah2024} 
offer an attractive alternative as they solely require the integration of the EoM, bypassing the simultaneous study of transversal dynamics,
which requires a certain understanding of
dynamical systems theory.

In this work, we circumvent the analysis of the tangent space by reinterpreting the hyperbolic separation of neighboring trajectories through distinct cancellation patterns in their time evolution when projected onto a reference grid formed, 
in the simplest case,
by two phase-space coordinates.
For this purpose,
we introduce Strong Trajectorial Ontological Differentiation (STOD), a measure that quantifies trajectory
\emph{divergence}
by directly comparing the orbits that solve the EoM by departing from a given set of ICs.
Unlike standard indicators -but similar to LD-,
STOD requires only the integration of ICs and does not necessitate the evolution of the linearized flow to analyze transversal dynamics.
Instead, 
by establishing a conceptual link between network science and physical dynamics, 
we reframe the question of instability: rather than measuring the metric distance between two ICs, we evaluate the similarity of their temporal evolutions. 

The application of this measure to a
(sufficiently dense) grid of ICs yields a dimensionless scalar field that reveals the underlying manifold structure.
Drawing from the framework of Ontological Differentiation (OD) and its recent applications in complex lexical and semantic networks \cite{GarciaCuadrillero2026}, this approach measures trajectory similarity over time.
While STOD provides a broad framework for trajectory comparison, this paper focuses on the {\FinSTOD} variant, which analyzes trajectories in reverse order from their final states to characterize the flow's structure.
This trajectory-based perspective aligns with recent advances in higher-order dynamics and temporal networks, where the history and ordering of interactions are recognized as fundamental to a system's identity \cite{Lambiotte2019, Wichmann2020}.

The introduced methodology, implemented through a component-wise cancellation process, allows for the detection of LCS while remaining independent of analytical derivatives. Consequently, it avoids reliance on Jacobians or specific velocity integrals, which can be numerically sensitive or unavailable in discrete datasets.
This new approach bears a conceptual resemblance to the study of symbolic itineraries in chaotic systems, where coarse-grained histories characterize dynamical complexity \cite{Lind1995, Bandt2002}.
As demonstrated below, {\FinSTOD} yield results in excellent agreement with three 
solidly grounded
tools (FTLE, FLI, and LD) for both autonomous and non-autonomous systems, effectively identifying invariant manifolds and LCS alike. Ultimately, {\FinSTOD} represents a fundamentally novel addition to the nonlinear dynamics toolbox.

The remainder of this paper is organized as follows. In Section~\ref{sec:definition}, we introduce STOD and the {\FinSTOD} variant. In Section~\ref{sec:methodology}, we describe the trajectory comparison framework and the standard indicators. Section~\ref{sec:systems} describes the suite of five popular
dynamical systems used for testing. Section~\ref{sec:results} presents a detailed comparison between {\FinSTOD} and FTLE, FLI and LD, highlighting its performance throughout different regimes. Finally, we conclude in Section~\ref{sec:conclusions} with a discussion on the implications of this ontological approach for the broader study of nonlinear dynamics.

\section{Strong Trajectorial Ontological Differentiation Definition}
\label{sec:definition}

STOD,
as well as the {\FinSTOD} version
mostly considered in this work,
quantifies the different dynamical behavior of the trajectories in a given system.
It is based on 
the OD framework~\cite{GarciaCuadrillero2026},
which was originally developed to measure semantic distance in lexical networks
by analyzing the recursive structure of word definitions.
In that context, a word's identity is defined by its dictionary definition,
i.\,e., by its composition in terms of other words of the same lexicon.

The fundamental premise of OD is that 
the semantic distance between any two concepts can be understood as a measure of the definitional transformation required for their underlying structures to align. OD operationalizes this by repeatedly expanding each initial concept into the terms that appear in its definition, producing successive layers
of words, in the case of lexical networks,
or
grid coordinates,
in the case of dynamical systems.
When a term is repeated during this process, a \emph{cancellation} occurs, signaling a point of semantic overlap or dynamical similarity,
depending on the case study.
The resulting distance value is then a function of the depth at which these cancellations manifest.

The framework allows for different variants based on specific cancellation rules.
In particular,
within Strong Ontological Differentiation (SOD),
lexical cancellations are triggered only by overlaps between elements originating from different starting concepts (cross-branch repetitions). This strict rule ensures that internal repetitions within a single concept's expansion do not, by themselves, signify semantic convergence with another concept.

In this work, we extend the OD framework to continuous dynamical systems
as a variant that we term \emph{Trajectorial Ontological Differentiation (TOD)}.
TOD differs from its lexical counterpart in the interpretation of the expansion process: instead of a recursive definitional expansion,
TOD employs a consequential chronological evolution.
The IC
$\mathbf{x}_0$
is (numerically) 
integrated
over (discrete) time steps
as prescribed by the EoM
up to a certain maximum time,
and the possible
different time evolutions
as a consequence of 
exponential sensitivity to the ICs
is measured by tracking the \emph{overlap}
of the trajectory
coordinates on a reference
(phase-space) grid.

More specifically,
we focus here on the strong variant of the mentioned iterative framework,
termed \emph{Strong Trajectorial Ontological Differentiation (STOD)}.
This novel OD version maintains the core principle of cross-cancellation from SOD but applies it to the discretized components
(both in time and phase-space)
of the trajectories.
By employing this strong variant, STOD reframes the question of trajectory separation
due to the exponential sensitivity to ICs that appears in chaotic motion: instead of a purely geometric measure of local separation, 
STOD provides a measure of ontological similarity by asking to what extent the composition of two (phase-space)
orbits corresponds to one another.

\subsection{Physical Motivation and Interpretation}
As it is well known,
chaotic motion is characterized by extreme sensitivity to the ICs,
which may yield large separations over finite times,
even for ICs that lie very close to each other (in phase space)~\cite{LL10}.
STOD efficiently quantifies this different dynamical behavior
of neighboring ICs.
Under regular motion,
neighboring trajectories are expected to share a high degree of similarity as their motion is restricted to the surface of invariant tori~\cite{Strogatz24},
resulting in low STOD values 
because the orbits generated 
sojourn in the same regions of the phase space.
Conversely,
under chaotic motion,
high STOD values are obtained due to 
the radically different historical fate.
When computed over a sufficiently dense grid of ICs,
STOD adequately identifies the phase-space regions with different dynamical behavior
and 
can also be used to identify the objects that determine the intricate chaotic dynamics,
such as the invariant manifolds or LCS.

\subsection{Formal Definition of STOD}
\label{sec:definitions}
In  this section,
we provide the definitions that form the STOD
cornerstone.

\begin{definition}[System Evolution and Discretization]
Consider a dynamical system described by
a set of ordinary differential equations $\dot{\mathbf{x}} = \mathbf{f}(\mathbf{x}, t)$
in an $n$-dimensional space $\mathcal{M} \subseteq \mathbb{R}^n$.
$\mathbf{f}(\mathbf{x}, t)$ provides the EoM of the system.
Let the trajectories originating from two ICs
$\mathbf{x}_A(0)$ and $\mathbf{x}_B(0)$ be defined by the sequences of discretized
$m$ grid coordinates:
$\mathcal{T}_A = \{ \mathbf{\C}_A(0), \mathbf{\C}_A(1), \dots, \mathbf{\C}_A(L) \}$
and
$\mathcal{T}_B = \{ \mathbf{\C}_B(0), \mathbf{\C}_B(1), \dots, \mathbf{\C}_B(L) \}$,
where
$\mathbf{\C}(k) = (\C_1(k), \dots, \C_m(k)) \in \mathbb{Z}^m$, with~$m \le n$,
represents the vector with the indices of the grid cell that is visited at
time~$t = k \, \Delta t$
being~$\Delta t$ the time step,
and $L$ is the total number of steps,
such that the total time equals~$\tau = L \,  \Delta t$.

Thus,
besides the usual time-discretized solution of the EoM,
STOD also requires a discretization of
the configuration and/or the phase space
in order to
allow for the comparison of 
the (discretized) orbits
{\TA} and {\TB}
using the component-wise cancellation logic
provided below
(see the worked example in Sec.~\ref{sec:example}).
Noticeably,
the dimension~$m$ of the created grid
must be smaller than or equal to the phase-space dimension~$n$.
However,
for the sake of clarity,
we limit this work to 2D grids ($m=2$),
where the grid coordinates correspond to two
phase-space coordinates.

\end{definition}


\begin{definition}[STOD Cancellation Rule]
A component value $\C_{Ai}(k)$ of trajectory {\TA} at level $k$ is marked as canceled if that same value has appeared for the corresponding component of the other trajectory, {\TB}, at any level $j$ up to the current level $k$ in its trajectory. Formally, $\C_{Ai}(k)$ is canceled if there exists $j \in [0, k]$ such that $\C_{Ai}(k) = \C_{Bi}(j)$. The same rule applies symmetrically for trajectory {\TB}.
\end{definition}

\begin{definition}[STOD Termination Rule]
The comparison process terminates at the first level $k^*$ where all components of any discretized vector visited by one trajectory up to that level have been canceled by the components visited by the other trajectory. That is, the process stops if there exists a level $j \in [0, k^*]$ such that for trajectory {\TA}, all components
$\{\C_{A1}(j), \dots, \C_{An}(j)\}$ 
are canceled, or if for trajectory {\TB}, all components
$\{\C_{B1}(j), \dots, \C_{Bn}(j)\}$
are canceled. If no such condition is met within the path length, the process terminates at the end of the available trajectory.
\end{definition}

\begin{definition}[STOD Value]
Once the comparison process terminates at level $k^*$, we determine the final state of all components visited up to that level. A component $\C_{Ai}(k)$
is marked as canceled if 
$\C_{Ai}(k) = \C_{Bi}(k')$ for~$k, k'  \le k^*$.
After resolving all cancellations, we count the number of uncanceled components on both trajectories at each level $k$, denoted as $U_k$. The total STOD value is the cumulative level-weighted sum of these uncanceled contributions as
\begin{equation}
    \text{STOD} = \sum_{k=0}^{k^*} k \cdot U_k.
    \label{eq:stod_value}
\end{equation}
\end{definition}

\subsection{{\FinSTOD}}
\label{sec:variants}
In this work, we focus on the {\FinSTOD} variant, 
which is defined as follows:

\begin{definition}[{\FinSTOD} Variant]

{\FinSTOD} computes the STOD value provided by~\eqref{eq:stod_value}
starting from the final position.
In other words,
instead of performing the comparison of the grid components
{\TA} 
and
{\TB} 
as provided in Definition~1,
it considers the
time-reversed sequences
$\mathcal{T}'_\alpha = \{ \mathbf{\C}_\alpha(L), \mathbf{\C}_\alpha(L-1), \dots, \mathbf{\C}_\alpha(0) \}$
being~$\alpha = A, B$.

\end{definition}

Consequently,
the level~0 of {\FinSTOD} 
corresponds to the final grid
position reached at the end of the integration window~$\tau$.
To facilitate the use and reproduction of this methodology, we have included the complete STOD and {\FinSTOD} logic in our public repository~\cite{FinSTOD_repo}.
Furthermore,
to facilitate the reader’s understanding of the procedure,
we present next a detailed worked example.

\subsubsection{Trajectory-pair Types}
\label{sec:types}
For any pair of 
ICs,
the comparison of their trajectories
{\TA} and~{\TB}
results in both a numerical STOD value and a classification into one of 
following three fundamental types~$T_{AB}$
within the integration time~$\tau$:

\begin{itemize}
    \item Type T (Terminated): The trajectories meet the termination condition (Definition 4).
    \item Type UC (Unterminated Cancelled): The trajectories do not meet the termination condition,
    but exhibit at least one instance of component cancellation.
    \item Type UU (Unterminated Uncancelled): The trajectories do not meet the termination condition,
    and exhibit zero instances of component cancellation.
\end{itemize}

This classification will be used to improve readibility of
the {\FinSTOD} scalar fields obtained over 2D grids of ICs
(see Sec.~\ref{sec:local_application_stod}).

\subsection{Worked Example: Serpentine Path}
\label{sec:example}
\begin{figure}[h!]
\centering
\includegraphics[width=0.5\columnwidth]{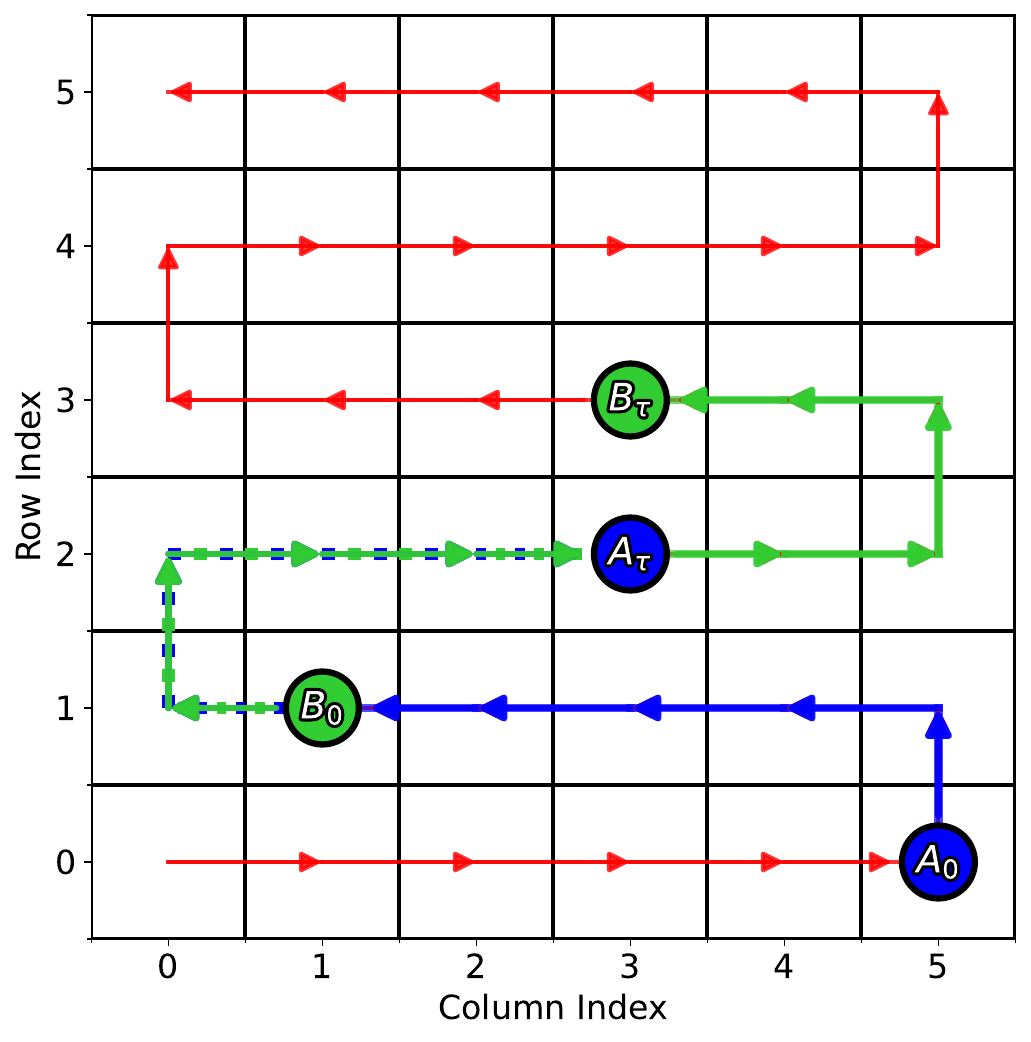} 
\caption{$6 \times 6$ grid showing the serpentine path (red). The trajectories originating from initial conditions $A_0$ (blue) and $B_0$ (green) are shown for a duration of $\tau=10$ time units, reaching final positions $A_\tau$ and $B_\tau$. Arrows indicate the direction of movement along the full path and the specific trajectories. The overlapping segment between both trajectories is shown as alternating blue and green dashes.}
\label{fig:serpentine_6x6}
\end{figure}

To illustrate the methodology, we apply {\FinSTOD} to
the $6\times6$ serpentine path shown in Fig.~\ref{fig:serpentine_6x6}.
Notice that the dynamical systems studied in the remaining of this work are defined by continuous EoM,
where
our analysis relies on their numerical integration and discretization onto 
high-resolution grids,
as detailed in Section~\ref{sec:methodology}.
Still, 
this simple and discrete example serves to clarify the component-wise cancellation logic within
STOD
framework.

We assume that the
trajectories {\TA} and {\TB}
evolve uniformly 
along the predefined path shown in Fig.~\ref{fig:serpentine_6x6},
which
starts at the position $(0,0)$ and then
moves alternatively to the right, upward and to the left
up to position~$(0, 5)$,
as sketched.
The position of each particle
a time~$t$ is represented as a 2D vector $\mathbf{\C}_{A/B, t} = (c_{A/B, t}, r_{A/B, t})$
corresponding, also, to the column and row grid indices, respectively.
We compare
trajectory~{\TA},
starting at the 
IC
$A_0$ [indices $(5, 0)$],
and
trajectory {\TB},
starting at $B_0$ [indices $(1, 1)$].
Their respective final states after $\tau = 10$ time units are denoted as
$A_\tau$ [indices  $(c_{A, \tau}, r_{A, \tau}) = (3, 2)$]
and $B_\tau$ 
[indices  $(c_{B, \tau}, r_{B, \tau}) = (3, 3)$].

The {\FinSTOD} value is calculated iteratively,
as summarized in Table~\ref{tab:FinSTOD_calculation}.
Canceled components of
trajectory~{\TA}
and the terms of 
trajectory~{\TB}
that are responsible for this cancellation
are marked with different dagger symbols.
The trajectories are compared level by level. At each step, we update the sets of visited grid components for each trajectory,
and check for new cancellations (Definition 3) and the termination condition (Definition 4).
To better understand the logic of the procedure,
we describe next the 
cancellation process step by step
as follows:
\begin{itemize}
    \item \textbf{Level 0:}
both trajectories are located at their final positions after
an evolution of $\tau = 10$ units of time.
At this point,
trajectory~{\TA} is found at 
$(c_{A, \tau}, r_{A, \tau})$,
while trajectory $B$ is located at
$(c_{B, \tau}, r_{B, \tau})$.
As the column component ($c_{A, \tau} = c_{B, \tau} = 3$) is the same,
it is canceled for both trajectories
(see terms marked with the dagger symbol~$\dag$
in Table~\ref{tab:FinSTOD_calculation}).
No state is fully canceled yet.
    \item \textbf{Level 1:}
    trajectory~{\TA}
    is at $(2, 2)$ and
    trajectory~{\TB} is at $(4, 3)$.
    Since the new position of 
    trajectory~{\TB}
    does not share any term with $(c_{A, \tau}, r_{A, \tau})$, so new cancellations occur.
    \item \textbf{Level 2:}
    trajectory~{\TA} is at $(1, 2)$ and
    trajectory~{\TB}
    is at $(5, 3)$.
As in the previous level,
no new cancellations occur.
    \item \textbf{Level 3:}
    trajectory~{\TA} moves to (0, 2),
    while trajectory~{\TB}
    reaches $(5, 2)$, introducing the row component~2.
    This value has already appeared in trajectory~$A$'s history (at levels 0, 1, 2, and 3),
so it is marked as canceled for trajectory~{\TA}
at those levels
(see terms highlighted with the dagger symbol~$\ddag$
in Table~\ref{tab:FinSTOD_calculation}).
Crucially, for 
trajectory~{\TA}
at level 0,
both the column ($c_{A, \tau} = 3$) and the row ($r_{A, \tau} = 2$) have now been matched by 
trajectory~{\TB}'s path.
Consequently,
since these two components of final position
are now fully canceled
(see Definition~3 above), the process terminates at $k^*=3$.
\end{itemize}

Once the termination condition is met,
we count the number of uncanceled components $U_k$ at each level based on the final state of cancellations
(see Table~\ref{tab:FinSTOD_calculation}).
As prescribed by \eqref{eq:stod_value},
the final {\FinSTOD} value is the sum of these 
level-weighted contributions: $0 + 3 + 6 + 6 = 15$.

According to the trajectory-pair types
presented in Sec.~\ref{sec:types},
the discussed trajectories
are of type~T only for evolution times~$\tau \ge 4$.
Smaller values of~$\tau$ would always lead to
a partial cancellation of~$c_{A,\tau}$
and, then, to a UC-type pair.

\begin{table}[h!]
\centering
\caption{Step-by-step calculation of {\FinSTOD} for the 
trajectories~{\TA} $(c_A, r_A)$ and~{\TB} $(c_B, r_B)$
that follow the $6\times6$ serpentine path shown in Fig.~\ref{fig:serpentine_6x6}.
Level 0 corresponds to the final discretized
position $(c, r)$ reached at the end of the time window,
where $r$ and $c$ denote the row and column indices, respectively.
The trajectories are compared in reverse order.
The process stops at level 3 because the termination condition (Definition~4)
has been met for the state of 
trajectory~{\TA}
at level 0,
a situation where the total number of uncanceled terms at level~$k$ ($U_k$) is calculated.
The canceling positions of 
trajectory~{\TB}
and
the canceled terms of 
trajectory~{\TA}
are highlighted with dagger symbols.}
\label{tab:FinSTOD_calculation}
\small
\begin{tabular}{|c|c|c|c|c|c|c|}
\hline
\textbf{Level ($k$)} &
$\mathbf{(c_A, r_A)}$ & $\mathbf{(c_B, r_B)}$ &
\textbf{Action} & $U_k$  & $k \cdot U_k$ 
\\
\hline
0 & $(3^\dag, 2^\ddag)$ & $(3^\dag, 3)$ &  Cancel position 3 ($\dag$) &  1& 0 \\ 
1 & $(2, 2^\ddag)$ & $(4, 3)$ & None & 3 & 3  \\ 
2 & $(1, 2^\ddag)$ & $(5, 3)$ & None &3 & 6 \\ 
3 & $(0, 2^\ddag)$ & $(5, 2^\ddag)$ &Cancel position 2 ($\ddag$) &  2 & 6 \\ 
\hline
\multicolumn{5}{|r|}{\textbf{{\FinSTOD}}} & \textbf{15} \\
\hline
\end{tabular}
\end{table}

This worked example demonstrates how the STOD scoring function~\eqref{eq:stod_value} penalizes 
trajectories
that remain distinct for longer time intervals,
which are associated with larger~$k$ values.
Although the serpentine path is a 1D constraint in a 2D  discretized
coordinate space (effectively a system with
1 DoF),
it provides a clear visualization of how component-wise cancellations (Definition 3) and the termination rule (Definition 4) operate. In the continuous systems analyzed in Section~\ref{sec:systems}, which possess more 
DoF
and complex structures in their respective spaces, the same logic is applied 
by performing a similar cancellation procedure in 
sufficiently dense 2D grids
formed by two phase-space coordinates of the
trajectories.

\section{Methodology and Application}
\label{sec:methodology}
To assess the utility of STOD as a tool for analyzing dynamical systems,
we apply it to a suite of 
paradigmatic
models.
When computed
over a dense grid of ICs,
the 
{\FinSTOD}
defines a scalar field, whose structure strongly resembles
that exerted by well-established chaos indicators
such as the FTLE, the FLI, and the  LD.
As described below,
this remarkable similarity will allow us to
efficiently identify invariant manifolds
and LCS.

\subsection{Standard Indicators}

In this section, we briefly review the standard indicators used as references for our study.
The excellent performance of the indicators considered
(FTLE, FLI and LD)
to identify phase-space structures
has been extensively illustrated in series of 
papers
within the fields of
molecular~\cite{Craven15, Revuelta19, Revuelta21, Revuelta23},
fluid~\cite{Haller2001, Shadden2005, Mancho2013, Haller2015, Koltai2018, Karrasch2020, Abdallah2024},
and
celestial mechanics~\cite{Froeschle1997, CG21, Guzzo2023, Flores26}, among others.

\subsubsection{Finite-Time Lyapunov Exponent Calculation}
The FTLE is a well-established measure of the maximal rate of separation of infinitesimally close 
ICs
over a finite-time interval~\cite{Haller2001, Haller2015}. It is computed from the flow map, $\mathbf{x}_{\tau} = \Phi_{t_0}^{t_0+\tau}(\mathbf{x}_0)$,
which maps an 
IC
$\mathbf{x}_0$ in the state space $\mathcal{M} \subseteq \mathbb{R}^n$ at time $t_0$ to its final state~$\mathbf{x}_\tau$ at time $t_0+\tau$.

The FTLE field, $\sigma(\mathbf{x}_0, T)$, is calculated as
\begin{equation}
    \sigma(\mathbf{x}_0, \tau) = \frac{1}{2 \vert \tau \vert} \ln(\lambda_{\rm max}),
    \label{eq:ftle}
\end{equation}
where $\lambda_{\rm max}$ is the maximum eigenvalue of the right Cauchy-Green strain tensor, $\mathbf{C}(\mathbf{x}_0, \tau)$.
This
tensor
is defined by the gradient of the flow map, $\mathbf{F} = \nabla \Phi_{t_0}^{t_0+\tau}(\mathbf{x}_0)$, as
\begin{equation}
    \mathbf{C}(\mathbf{x}_0, \tau) = \mathbf{F}^T \mathbf{F}.
    \label{eq:cauchy_green}
\end{equation}
High values in the FTLE field (ridges) reveal lines of intense stretching that act as transport barriers, a result that allows the proper identification of the LCS
that are responsible for, e.\,g.,
the transport dynamics in oceanographic systems~\cite{Haller2015}.

\subsubsection{Fast Lyapunov Indicator Calculation}

The FLI~\cite{Froeschle1997, Guzzo2023} is another popular chaos indicator that measures the growth of a tangent vector $\mathbf{v}(t)$ 
as a trajectory evolves.
This vector belongs to the
tangent space $T_{\mathbf{x}(t)}\mathcal{M} \cong \mathbb{R}^n$
at each point $\mathbf{x}(t)$ of the trajectory.
The FLI is defined as
\begin{equation}
    \textnormal{FLI}(\mathbf{x}_0, \tau) = \sup_{0 \le t \le \tau} \ln \| \mathbf{v}(t) \|,
    \label{eq:fli}
\end{equation}
where $\| \cdot \|$ corresponds to the Euclidean norm 
of the vector $\mathbf{v}(t) = \mathbf{F}(t)\mathbf{v}(0)$,
which evolves according to the linearized dynamics.

Similar to the FTLE, the FLI identifies 
LCS
as regions of high values, but it provides better contrast in chaotic regions due to its focus on the maximum growth encountered along the orbit.

\subsubsection{Lagrangian Descriptors Calculation}
Unlike the previous methods,
which are grounded on the dynamics of the tangent space,
LD is an integral-based tool that requires only the analysis of the trajectory evolution.
In that sense,
it is a simpler tool since it does not require any notion of tangent space or linearized dynamics.
In its original definition~\cite{Mancho2013}, 
the LD were defined as 
\begin{equation}
    \textnormal{LD}(\mathbf{x}_0, \tau) = \int_{t_0}^{t_0+\tau} \| \mathbf{\dot x}(\mathbf{x}(t), t) \| dt,
    \label{eq:ld}
\end{equation}
which equals the arc length of the trajectory.
More advanced definitions have been proposed, including the use of $p$-norms~\cite{Lopesino2017, Revuelta19} and action-based formulations where the integrand is replaced by the Lagrangian of the system~\cite{Abdallah2024}.
LD reveals Lagrangian coherent structures as singular features (ridges or valleys) in the scalar field.

\subsection{Local Application of the STOD Framework}
\label{sec:local_application_stod}
The STOD framework provides a general methodology for trajectory comparison that can be implemented through different variants, such as the {\FinSTOD}
provided by the Definition~5 of Sec.~\ref{sec:definitions}.
To analyze the continuous dynamical systems studied in Section~\ref{sec:systems}, we must translate the pairwise comparison logic into a scalar field that can be visualized and compared to standard indicators. This is achieved through a local comparison scheme that assigns a single value to each initial condition based on its similarity to its immediate neighbors. 

It is important to emphasize that this local application methodology is universal within the STOD framework: any variant, including the {\FinSTOD} 
version considered
 in this work, is applied to a grid of ICs following the same selection and normalization logic. By describing this process in general terms, we provide a standard procedure for generating scalar fields from any  
ontological comparison. In our specific application, we consider a  2D grid of 
\emph{cells} of ICs  $\C_{i}$
 and compute the pairwise {\FinSTOD} values between each point and its four nearest neighbors in the von Neumann neighborhood.

\subsubsection{Selection Logic and Scalar-Field Generation}
For a given system defined on a grid of ICs,
the scalar value assigned to a central cell $\C_i$
is determined through a selection process within its local neighborhood that identifies the highest degree of 
dynamical \emph{dissimilarity}:

\begin{enumerate}
    \item Identify the four nearest cells of $\C_i$ 
    (its von Neumann neighborhood):
    $\left\{\C_{\textnormal{up}}, \C_{\textnormal{down}}, \right.$
    $\left. \C_{\textnormal{left}}, \C_{\textnormal{right}}\right\}$.

    \item Compute the pairwise {\FinSTOD}
    (or any other STOD version)
 value 
 and type classification ($T_{ij}$)
 as described in Sec.~\ref{sec:types}
between the trajectory of $\C_i$ and each von Neumann neighbor~$\C_j$.

    \item Assign to cell $\C_i$ the {\FinSTOD} value
               whose associated class has the highest priority,
               following the order UU $>$ UC $>$ T, and,
               within that class, the largest numerical {\FinSTOD} value.
               This hierarchy reflects the degree of dynamical dissimilarity: 
               UU pairs exhibit no shared components whatsoever, indicating maximal divergence;
               UC pairs show partial overlap through cancellations but never reach termination;
               and T pairs demonstrate the highest similarity, as the termination condition implies
               a factual sharing of trajectory elements.
               The priority ordering ensures that, when different pair types yield comparable
               numerical scores at the same level, the classification reflecting greater
               dissimilarity prevails—analogous to how FTLE and FLI select the maximum
               stretching direction to characterize local dynamics.
    
\end{enumerate}

Once we have the types for the whole grid,
we compute a proportion $P$ to add an additional normalization to
 the value {\FinSTOD} at each unit cell~$\C_i$ as
\begin{equation}
    P = \frac{N_T}{N_T + N_{UC}},
    \label{eq:proportion}
\end{equation}
where $N_T$ and $N_{UC}$ are the counts of cells classified as 
type T and type UC, respectively. 
this value $P$ acts as a pivot for a segmented mapping that defines the
final  scalar field by scaling the {\FinSTOD} value in each cell as:
\begin{itemize}
    \item For T-type cells, the raw STOD values are normalized to the interval $[0, P]$.
    \item For UC-type  cells, the raw STOD values are normalized to the interval $[P, 0.99]$.
    \item For UU-type cells, a constant value of 1.0 is assigned.
\end{itemize}
This approach ensures that {\FinSTOD} highlights the boundaries between dynamically distinct regions, where a high value indicates a point of greater differences in the evolution of neighboring trajectories. 

Finally, we address the computational performance of the STOD framework. To evaluate this, we conducted a performance test using an Intel(R) Xeon(R) Gold 6230 CPU @ 2.10GHz. The core ontological comparison logic of {\FinSTOD} required an average of 2.70~s per $1000 \times 1000$ grid, while the standard indicators required an average of 9.54~s. However, the total STOD pipeline, which includes the numerical integration and storage of full trajectories, was approximately 2.9 times slower than the other methods, with the integration phase accounting for 89.0\% of the total time. These results indicate that while the requirement to store full trajectories introduces a necessary computational overhead, the STOD comparison itself is computed efficiently compared to standard indicators.

\section{Dynamical Systems Under Study}
\label{sec:systems}
We apply the standard indicators and the STOD framework -specifically the {\FinSTOD} variant- to five distinct dynamical systems, chosen to span a range of behaviors from systems with simple linear EoM,
to complex, time-dependent, and chaotic flows. By selecting systems with increasing levels of dynamical complexity, we effectively test the {\FinSTOD} indicator across a significant variety of regimes, from steady flows with analytic solutions to high-dimensional chaotic
attractors~\cite{Guckenheimer02, Moon1987}.
For each system, we define a high-resolution grid of
ICs;
this resolution is essential to ensure that the local, neighbor-based comparisons performed by {\FinSTOD} are meaningful and can capture fine structural details. The numerical integration of trajectories is performed using the LSODA algorithm (as implemented in the \texttt{scipy.integrate.odeint} package), which is a high-order, adaptive-step solver capable of handling both stiff and non-stiff systems~\cite{Petzold1983, Hindmarsh1983},
ensuring the accuracy of the long-term path identities required
for both the reference calculations provided
by FTLE, FLI and LD,
and the {\FinSTOD} evaluation.

\subsection{Autonomous systems}
\label{sec:systems_autonomous}

\subsubsection{Linear Hyperbolic Saddle Point}

The first case study corresponds to the linearized motion 
of a 1-DoF saddle point.
This is the simplest autonomous system
showing exponential sensitivity to the ICs.
The corresponding EoM
are given by
\begin{align}
    \frac{dx}{dt} &= \lambda x, \label{eq:linear_saddle_x} \\
    \frac{dy}{dt} &= -\lambda y, \label{eq:linear_saddle_y}
\end{align}
and their solutions 
by
\begin{align}
    x(t) &= x_0 e^{\lambda t},  \label{eq:linear_saddle_x2}  \\
    y(t) &= y_0 e^{-\lambda t}. \label{eq:linear_saddle_y2} 
\end{align}

\begin{figure}[ht!]
    \centering
    \includegraphics[width=0.5\columnwidth]{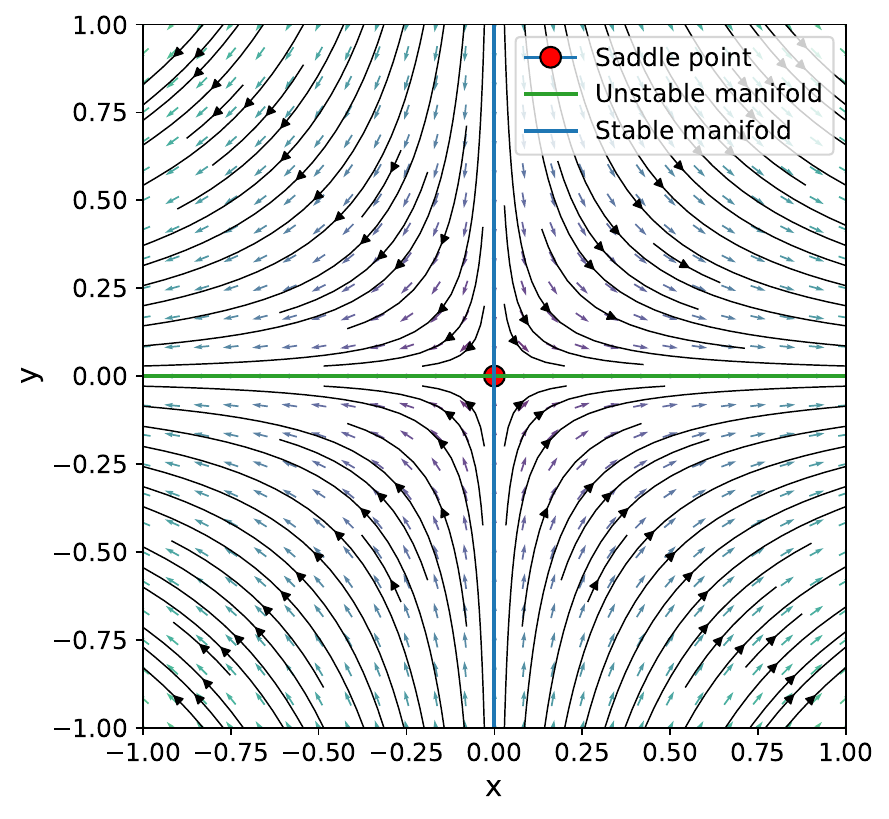}
    \caption{Vector field and invariant manifolds for the linear hyperbolic saddle point~\eqref{eq:linear_saddle_x}--\eqref{eq:linear_saddle_y} for~$\lambda=0.5$.
}
    \label{fig:linear_vectorfield}
\end{figure}

As shown in Fig.~\ref{fig:linear_vectorfield},
the  vector field 
is formed by a set of hyperbolas around the origin.
The time evolution \eqref{eq:linear_saddle_x2}-\eqref{eq:linear_saddle_y2}
of a generic  IC
approaches towards the origin in the direction of the $y$ axis
and separates from it in the direction of the $x$ axis
at rates that depend on the value of the $\lambda$ parameter.
These axes coincide with the invariant manifolds
with determine the dynamics of the system.
If the IC coincides with the origin, it remains there
for all times
 as it is an (unstable)  fixed point.


\subsubsection{Simple Pendulum}
\label{sec:pendulum}
The second system considered is a benchmark for regular motion: the simple pendulum.
The EoM of this autonomous 1-DoF system are given by
\begin{align}
    \frac{d\theta}{dt} &= \omega, \label{eq:pendulum_theta} \\
    \frac{d\omega}{dt} &= -\mu \sin(\theta), \label{eq:pendulum_omega}
\end{align}
where $\theta$ is the angle  that the pendulum forms with the vertical,
$\omega$ is the conjugate momentum,
and
$\mu = g/L$
equals the ratio between the gravitational acceleration~$g$ and the pendulum length~$L$.

As shown in Fig.~\ref{fig:pendulum_forw_raw},
due to the rotational symmetry,
the fundamental domain is defined within a $\theta$ range of~$2 \pi$\,rad,
e.\,g.,
$-\pi \le \theta \le \pi$.
The system has two (infinite) families of fixed points.
On the one hand,
the pairs
$(\theta_{\rm stab}(n), \omega_{\rm stab}) = (2n\pi, 0)$,
with~$n \in \mathbb{Z}$,
correspond to the elliptic point or center,
where the pendulum remains in equilibrium at its bottom position.
On the other hand,
$(\theta_{\rm unstab}(n), \omega_{\rm unstab}) =((2 n+1)\pi, 0)$
provides the unstable (hyperbolic) fixed point,
where the pendulum is located at its highest position.
The infinite series of $((2n+1)\pi, 0)$ points are connected
by separatrices that partition the phase space into regions of
libration, where oscillatory motion around the stable centers takes place, 
and rotations, where continuous circulation occurs. 
Remarkably,
the motion in the vicinity of the unstable fixed points is analogous to that observed 
around the linear hyperbolic saddle described in the previous section,
with the unstable (stable) manifold emerging from 
the point $((2n+1)\pi, 0)$
entering 
to the point $((2n+3)\pi, 0)$ as the
stable (unstable) manifold.
Physically, these manifolds describe the case where a slight distortion knocks the pendulum out of equilibrium when it is initially 
at its vertically position. The pendulum accelerates to a maximum velocity at~$\theta = 0$\,rad,
and subsequently slows down as it approaches the equilibrium point again,
reaching it only at infinite time.


\subsubsection{Lorenz'63 System}
\label{sec:lorenz}

The third system studied is also autonomous but it has 3 dimensions.
Originally derived to model 
the convection~\cite{Lorenz1963}
of a fluid is heated from below,
the Lorenz system is a 
paradigmatic example of deterministic chaos~\cite{Guckenheimer1983, Froyland09}.
Unlike the 2D systems previously discussed, the higher dimensionality of the Lorenz system allows for the emergence of a strange attractor~\cite{Sparrow1982, Tucker99}.
The  EoM are given by
\begin{align}
    \frac{dx}{dt} &= \sigma(y - x), \label{eq:lorenz_x} \\
    \frac{dy}{dt} &= x(\rho - z) - y, \label{eq:lorenz_y} \\
    \frac{dz}{dt} &= xy - \beta z. \label{eq:lorenz_z}
\end{align}
Physically,
the coordinates~$x$,~$y$, and~$z$ are related to
the intensity of convection,
the temperature difference between the rising and lowering currents,
and the deviation from linear behavior of the vertical temperature profile, 
respectively.
Moreover,~$\sigma$ corresponds to Prandtl number,
$\rho$ to the Rayleigh number,
and~$\beta$ is related to the physical dimensions of the fluid layer.

The enduring popularity of the Lorenz'63 model stems from its applicability far beyond atmospheric convection; it serves as a foundational framework for understanding strange attractors in diverse fields~\cite{Strogatz24}, including
laser dynamics,
electrical circuits,
and
chemical kinetics.
In those cases,
the parameters~$\sigma$,~$\rho$, and~$\beta$
have other physical meanings,
and their precise values are responsible for very different different dynamics.
For example,
for~$\rho<1$,
the origin is a 
hyperbolic sink and the only attractor~\cite{Guckenheimer1983}.
Thus, 
al trajectories eventually reached this point, 
no matter their IC.
For~$1< \rho< \rho_{\rm HB} = 24.74$,
the origin becomes unstable
and two stable points emerge due to a pitchfork bifurcation~\cite{Strogatz24}.
Within this regime,
all orbits tend to approximate one of these two points.
Larger values of~$\rho$ that surpass the~$\rho_{\rm HB}$
value of  a subcritical Hopf bifurcation~\cite{Strogatz24} make the dynamics dramatically much more involved
by opening the route for chaotic motion
with the appearance of a strange attractor
(though regular motion can still be observed under certain ICs)~\cite{Sparrow1982, Tucker99}.


\subsection{Non-autonomous systems}
\label{sec:systems_non-autonomous}
In order to assess the ability of the
{\FinSTOD}
to identify not only
invariant manifolds in autonomous systems as the ones discussed so far,
but also LCS in non-autonomous systems,
we present here the two additional benchmarks considered in this work:
the forced Duffing oscillator and the double gyre.

\subsubsection{Forced Duffing Oscillator}

The fourth system considered is
the Duffing equation,
which models a damped, driven oscillator with a nonlinear restoring force~\cite{Duffing1918}.
As Lorenz'63 (see Sec.~\ref{sec:lorenz}),
this system exhibits a strange attractor,
leading to strongly chaotic dynamics.
For a unit-mass oscillator,
the EoM are given by
\begin{align}
    \frac{dx}{dt} &= v, \label{eq:duffing_x} \\
    \frac{dv}{dt} &= \gamma \cos(\Omega t) - \delta v - \alpha x - \beta x^3, \label{eq:duffing_v}
\end{align}
where~$x$ gives the position (angle),
$v$ the (angular) velocity,
$\gamma$ the strength of the external modulation,
$\Omega$ its frequency,
$\delta$ is the friction,
$\alpha$ is the linear stiffness,
and
$\beta$ is the non-linear stiffness.
Notice that for~$\gamma = 0$
the system is autonomous
and equivalent to a conservative
quartic Hamiltonian~\cite{Guckenheimer1983}.
In this case,
for~$\alpha, \beta > 0$,
the origin is an equilibrium point
(a stable focus for~$\delta > 0$ or a center for $\delta = 0$).
For~$\alpha < 0 < \beta$,
the system resembles a double well
with a saddle point at the origin,
and two stable foci (for $\delta > 0$) located at
$\pm \sqrt{- \alpha / \beta}$.
The motion is noticeably modified by inclusion of an external forcing ($\gamma > 0$).
Then,
for low values of the force amplitude~$\gamma$,
the motion is regular and
takes place around one of the previous centers
or
following a limit cycle.
As the value of~$\gamma$ is increased,
cascades of bifurcations take place,
noticeably changing the structure of the phase space
with the appearance of an strange attractor that 
enables
highly chaotic motion.


\subsubsection{Time-Dependent Double Gyre}

The
double gyre is a 2D non-autonomous system 
that is routinely used in the study of chaotic mixing in time-periodic flows~\cite{Froyland09}.
It consists of two counter-rotating gyres whose boundary oscillates,
creating a chaotic mixing region \cite{Ottino1989}.
The EoM are given by
\begin{align}
    u(x,y,t) &= -\pi A \sin(\pi f) \cos(\pi y), \label{eq:dg_u} \\
    v(x,y,t) &=  \pi A \cos(\pi f) \sin(\pi y) \frac{\partial f}{\partial x}, \label{eq:dg_v}
\end{align}
where $f(x,t) = a(t)x^2 + b(t)x$, 
being $a(t) = \epsilon \sin(\kappa t)$ and $b(t) = 1 - 2a(t)$,
and
with $A$ accounting for the effects of the gyres,
$\epsilon$ the oscillation amplitude of the vertical boundary,
and 
$\kappa$
the corresponding angular frequency.
The double gyre is autonomous for~$\epsilon=0$.
In this case, the motion is completely regular around one of the two gyre centers,
which are separated by an invariant manifold that acts as a true separatrix
that crosses the origin (saddle point).
This separatrix corresponds to a heteroclinic connection
that joins the hyperbolic points $(1,0)$ and $(1,1)$~\cite{Shadden2005}.
In accordance with the KAM theorem,
the motion remains regular for most of the ICs 
for slightly larger values of~$\epsilon$,
but a thin chaos layer emerges around the saddle points.
Larger values of~$\epsilon$ break the separatrix 
by creation of a 
heteroclinic
tangle that is responsible for
chaotic motion.
Very low oscillations, which correspond to small $\kappa$ values,
only perturb slightly the regular motion;
in this case,
the flow is quasi-steady and the mixing degree low.
Under very fast oscillations,
which occur for small $\kappa$ values,
a very thin chaotic layer appears,
but regular motion still dominates.
A more interesting regimen is found for intermediate $\kappa$ values,
where a strong resonant coupling between the autonomous rotation frequency
and the time-periodic modulation.

\section{Results}
\label{sec:results}

In this section, we present the comparative results for the five
systems described in Sec.~\ref{sec:systems}.
We use a duration $\tau$ for the evolution of the systems, considering both forward and backward time-evolutions. For convenience, {\FinSTOD} results are shown in logarithmic scale ($\log(1+s)$), where $s$ represents the {\FinSTOD} value, for the complex systems where it provides better structural contrast. For a more comprehensive exploration, we have created a repository~\cite{FinSTOD_repo} where, besides the code for reproduction of {\FinSTOD}, one can also see the videos of the whole evolution comparison and both the raw and log evolution of {\FinSTOD} for all the systems studied.

Unveiling the hyperbolic nature of chaotic motion and its sensitivity to 
ICs
requires a sufficiently long
time window to identify invariant manifolds and LCS.
It is well-known that dynamical systems require a characteristic minimum integration time necessary to unravel their underlying \emph{skeleton}.
When trajectories have less time to build up their dynamics, the indicators struggle to extract relevant information, resulting in relatively featureless or sparse scalar fields. This duration is typically determined by the characteristic timescales of the system, such as the Lyapunov time or the period of the external driving forces.

In what follows,
traditional indicators (FTLE, FLI, and LD) are displayed using their numerical values, with LD shown in logarithmic scale for specific cases, such as the linear saddle, where this transformation is known to improve the resolution of the manifold structures. For most other systems, we primarily show the raw values of the standard indicators to avoid over-homogenization of the resulting scalar fields.
All scalar fields are min-max normalized to $[0, 1]$ for visual consistency.
Recall that rather than the precise value of each chaos indicator,
what matters are the differences between neighboring ICs as they allow the identifation of
the phase-space structures that determine the system dynamics.
We put the color scale bar on the first plot (Fig.~\ref{fig:linear_forw_static}) so one can see what the reference is; in this scale bar, we have also put the trajectory classification for that case so one can understand how the proportion $P$ partitions the diagnostic range. Of course, this classification and the specific value of $P$ would change from system to system.

\subsection{Autonomous systems}
This section is devoted to discuss the 
{\FinSTOD} performance for the autonomous systems
introduced in Sec.~\ref{sec:systems_autonomous}.
\subsubsection{Linear Hyperbolic Saddle Point}
We begin our analysis with the linear hyperbolic saddle described by
\eqref{eq:linear_saddle_x}--\eqref{eq:linear_saddle_y2}
for~$\lambda = 0.5$ and~$\tau = 15$.
The computations are performed on a $1000 \times 1000$ grid of
ICs
spanning
the domain $x \in [-10, 10]$, $y \in [-10, 10]$.
In this system,  
other tools like the FTLE and FLI
may be less informative as they produce a constant field~\cite{Haller2011, Farazmand12}. Because the Jacobian of the linear velocity field is spatially constant, these indicators result in uniform scalar fields across the entire domain, failing to resolve the underlying manifold structure.

\begin{figure}[ht!]
    \centering
    \includegraphics[width=0.8\textwidth]{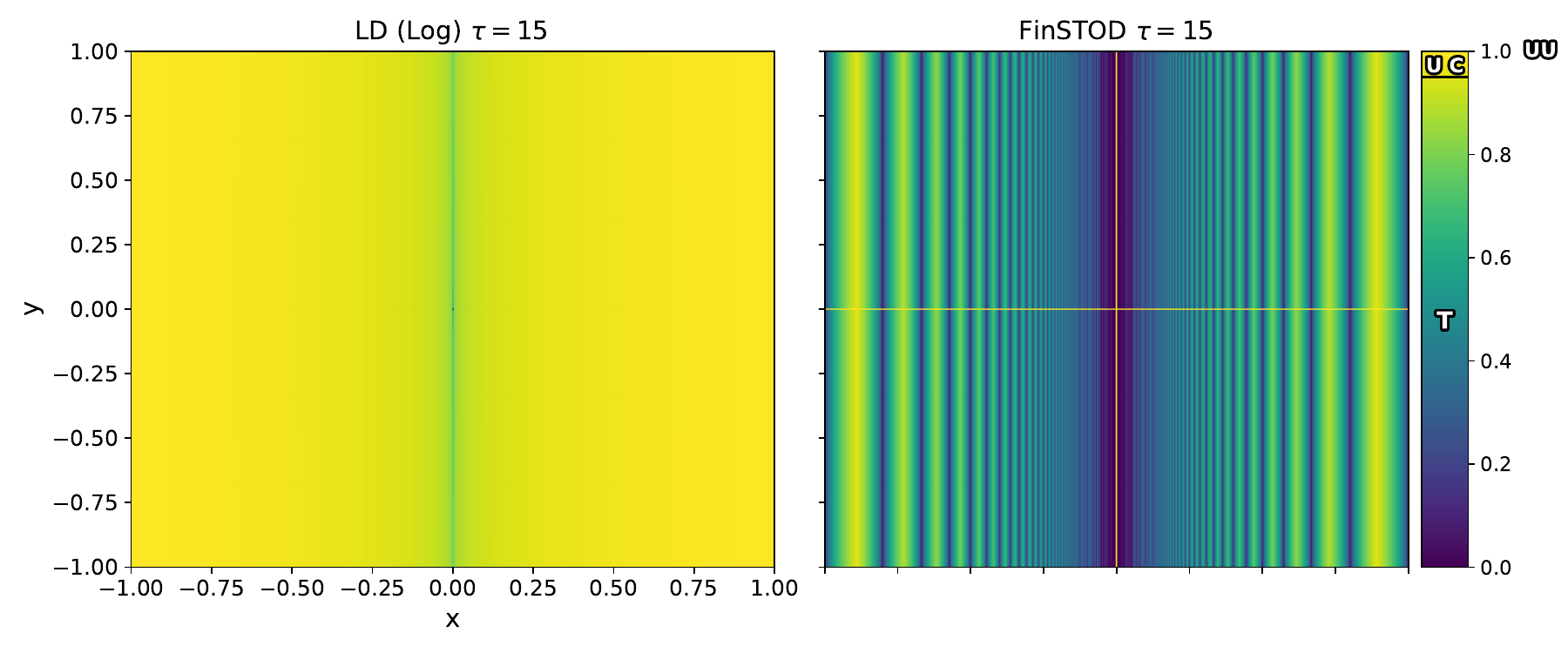}
    \caption{(Left) Lagrangian descriptors~\eqref{eq:ld} in logarithmic scale
                      and (right) {\FinSTOD}~\eqref{eq:stod_value} for the linear saddle~\eqref{eq:linear_saddle_x}-\eqref{eq:linear_saddle_y}
                      computed forward in time for a duration $\tau=15$.}
    \label{fig:linear_forw_static}
\end{figure}

\begin{figure}[ht!]
    \centering
    \includegraphics[width=0.8\textwidth]{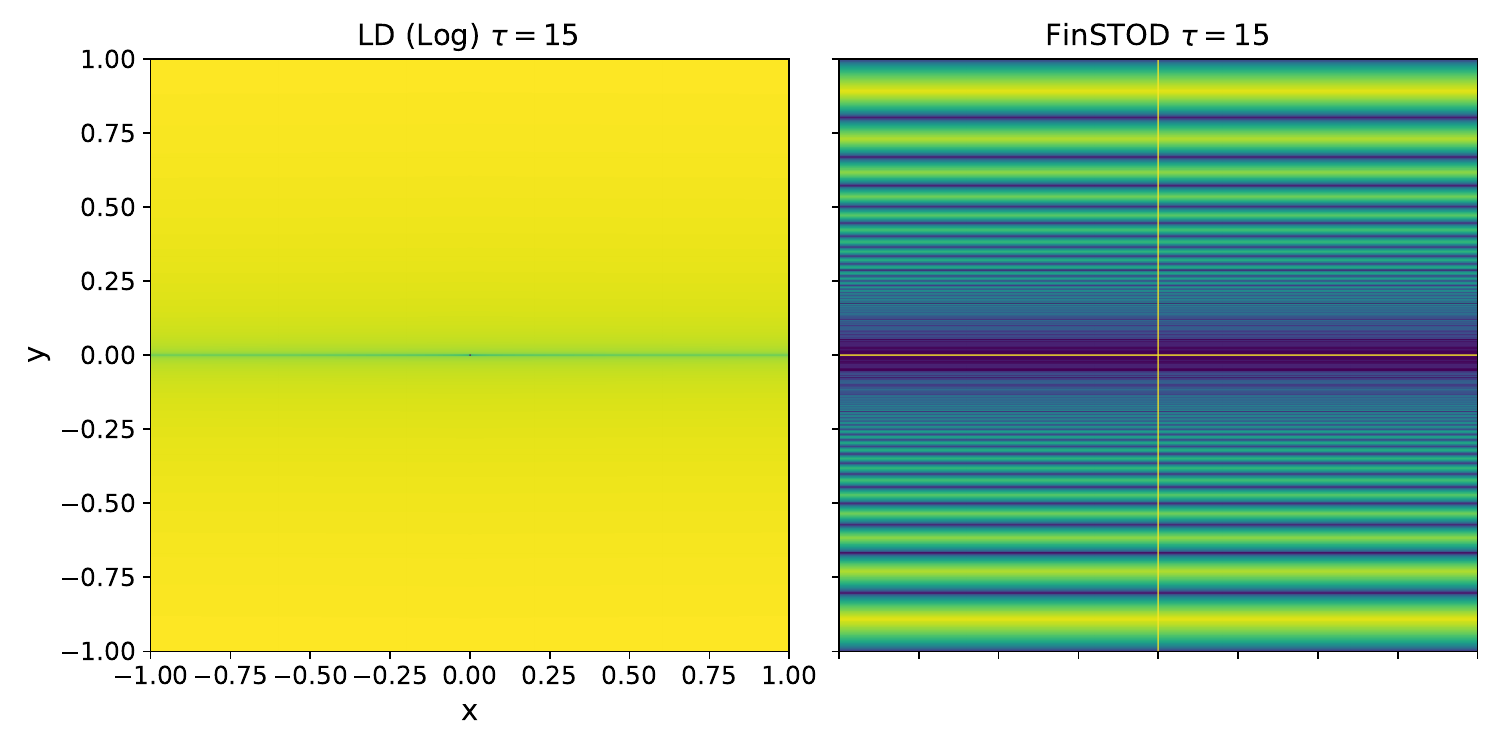}
    \caption{Same as Fig.~\ref{fig:linear_forw_static} for the backward time evolution.}
    \label{fig:linear_back_static}
\end{figure}

Consequently, we compare the {\FinSTOD} solely with the LD.
Figures~\ref{fig:linear_forw_static} and \ref{fig:linear_back_static} show the results for forward and backward time-evolutions,
respectively.
Notice that the forward evolution of {\FinSTOD}
adequately identifies 
not only the stable manifold,
as the LD does,
that corresponds to the vertical axis of Fig.~\ref{fig:linear_forw_static},
but also, in slightly less precise fashion,
the unstable manifold given by the horizontal axis.
Similarly,
the backward time evolution unravels
both manifolds,
while the LD only identify the unstable one,
which coincides with the horizontal axis of Fig.~\ref{fig:linear_back_static}.

The results demonstrate that {\FinSTOD}
successfully identifies both the stable and unstable manifolds,
appearing as a clear cross-shape
centered at the origin. In the forward-time analysis (Fig.~\ref{fig:linear_forw_static}), {\FinSTOD} resolves the structure with high contrast, showing a decaying density of blue (low separation) along the manifolds as the duration increases. This behavior is qualitatively similar to the LD results, but {\FinSTOD} provides a sharper identification of the manifold intersection. At long durations, LD tends to show one dominant axis, whereas {\FinSTOD} maintains the visibility of both the stable and unstable manifolds simultaneously, providing a more complete picture of the saddle's influence on the local flow. This occurs because, for a sufficiently long integration time, all ICs approach the unstable (stable) manifold when evolved forward (backward) in time, except for those exactly on the stable (unstable) manifold.

A more detailed exploration of the influence of integration time on the resulting scalar fields is provided in our repository~\cite{FinSTOD_repo}, where we include
the \texttt{A\_linear\_saddle} video series showing the continuous time evolution up to the shown duration, allowing for a direct comparison of how the manifold structures emerge and stabilize in both {\FinSTOD} and LD.

\subsubsection{Simple Pendulum}
\begin{figure}[ht!]
    \centering
    \includegraphics[width=0.8\textwidth]{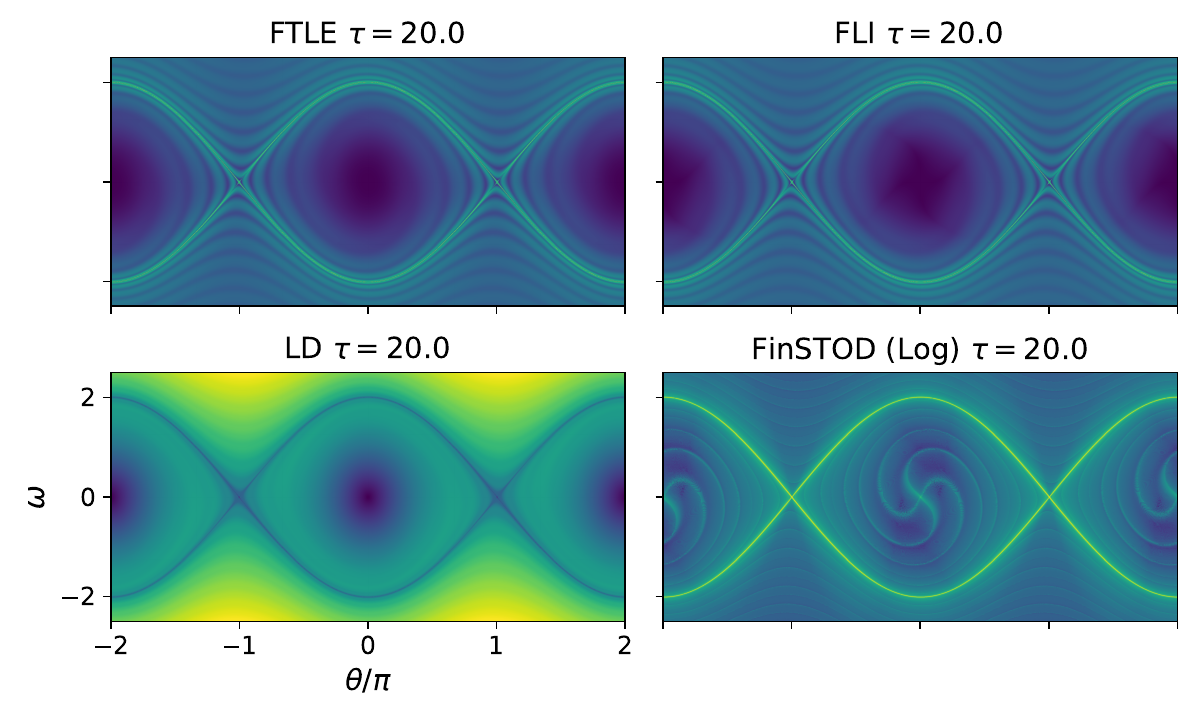}
    \caption{Comparative analysis for the simple pendulum
    \eqref{eq:pendulum_theta}-\eqref{eq:pendulum_omega} at duration $\tau=20$ (Forward).
     From top-left to bottom-right:
     Finite-time Lyapunov exponent (FTLE)~\eqref{eq:ftle},
     fast Lyapunov indicator (FLI)~\eqref{eq:fli},
     Lagrangian descriptors (LD)~\eqref{eq:ld},
     and {\FinSTOD}~\eqref{eq:stod_value}.}
    \label{fig:pendulum_forw_raw}
\end{figure}

\begin{figure}[ht!]
    \centering
    \includegraphics[width=0.8\textwidth]{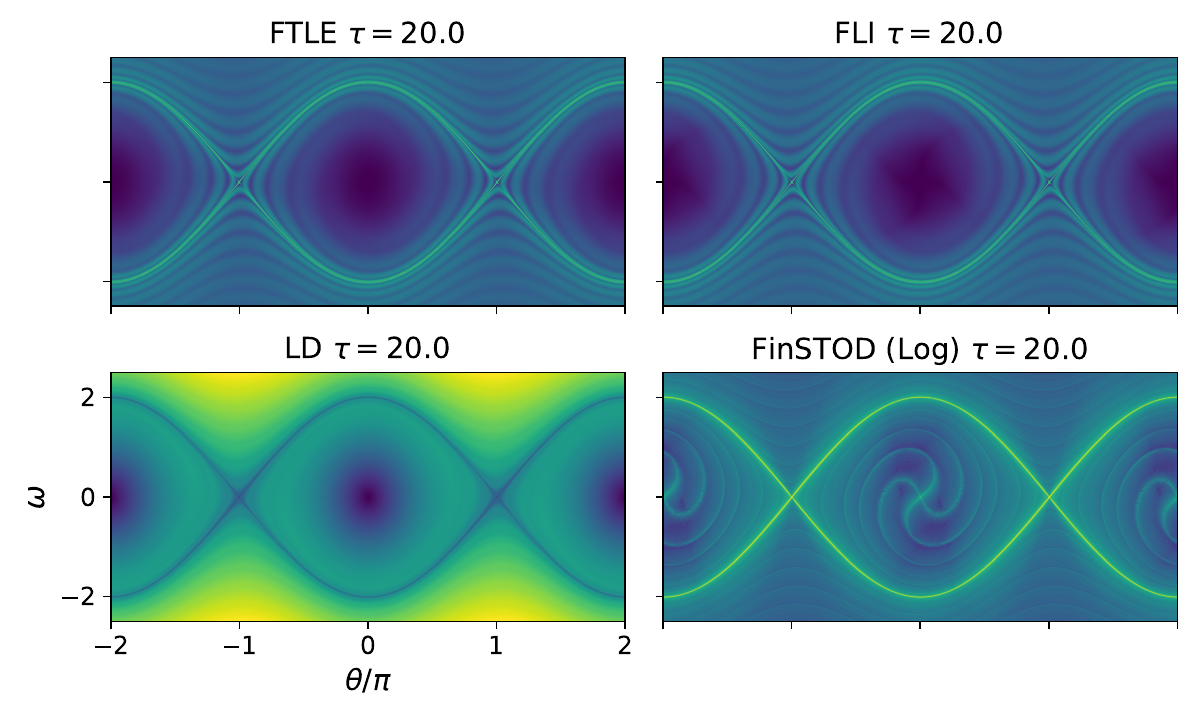}
    \caption{Same as Fig.~\ref{fig:pendulum_forw_raw} for a backwards time evolution.}
    \label{fig:pendulum_back_raw}
\end{figure}

The results for the simple pendulum with~$\mu = 1$ and~$\tau = 1$,
governed by 
\eqref{eq:pendulum_theta} and \eqref{eq:pendulum_omega}, are presented in Figs.~\ref{fig:pendulum_forw_raw} and \ref{fig:pendulum_back_raw}
for the forward and backward evolutions, respectively.
In order to highlight the periodic structure of the phase space,
the analysis is performed on a $1000 \times 1000$ grid of 
ICs
defined over
two fundamental domains $\theta \in [-2\pi, 2\pi]$, $\omega \in [-2.5, 2.5]$.

First, notice that
all considered indicators present the same structure,
with the invariant manifolds emanating from the unstable fixed point(s)
located at~$\theta_{\rm unstab}(-1)=- \pi$\,rad
and~$\theta_{\rm unstab}(0) = \pi/2$\,rad
(see Sec.~\ref{sec:pendulum}).
Remarkably,
while the separatrices, i.\,e., 
the invariant manifolds,
are highlighted with maximum values (ridges)
for the FTLE, the FLI, and the {\FinSTOD},
they correspond to minima in the case of the LD
(they are indeed singularities,
where the derivative diverges~\cite{Lopesino17}).
Noticeably,
these structures are more clearly identified for {\FinSTOD}
compared to the rest of the indicators,
as they have much larger values over  the manifolds
than in the rest of the phase-space points.

Second,
the mentioned invariant manifolds partition the phase space
into two regions with very different dynamical behavior.
On the one hand,
the outmost region contains a series of ripples
which are associated with the 
periodic rotational orbits around the origin.
These ripples are clearly visible in the plots of the FTLE, 
the FLI and the {\FinSTOD},
and less noticeable in the case of the LD.
On the other hand,
the inner region of the manifolds correspond to 
librational periodic orbits,
where the pendulum oscillates around
the central elliptic point~$\theta_{\rm stab}(n)$
(see Sec.~\ref{sec:pendulum}).
In this case,
the LD-plot is a slow varying function,
where no structure can be identified,
while in case of the rest of indicators 
spirals centered at the elliptic point can be observed,
specially in the {\FinSTOD} case.
Notice, nevertheless,
that the indicators considered have much smaller values along
the spirals than over the manifolds,
specially in the case of {\FinSTOD},
where a  logarithmic scale has been used.

Third,
the spirals emanating from the central elliptic point(s)~$\theta_{\rm stab}(n)$
have opposite orientations for
the backward-time evolution shown in Fig.~\ref{fig:pendulum_back_raw}
than in the forward-time case introduced in Fig.~\ref{fig:pendulum_forw_raw},
a result that is a consequence of the different dynamics.
The appearance of these spirals with time
can be observed in more detail in the
\CB{\texttt{B\_pendulum\_*}
repository videos~\cite{FinSTOD_repo}.}
Furthermore,
the manifolds that act as true phase-space separatrices are
equally well identified in both cases.
However,
as in the linear-saddle case,
the information that allows the identification of the structures
is not incorporated in the same way for these two time evolutions.
The 
forward-in-time evolution encodes the information of the invariant manifolds 
mostly along the stable direction of the manifold that emanates from the hyperbolic point
located at~$\theta_{\rm unstab}(n)$,
which are is given by the corresponding eigenvector~$(\theta, \omega) = \pm (1, - \sqrt \mu) = \pm ( 1, - 1)$,
while the backward evolution in time
greatly captures the information along the unstable direction
provided by the eigenvector~$(\theta, \omega) = \pm ( 1, \sqrt \mu) = \pm ( 1,  1)$.
For sufficiently long integration times ($\tau$),
a larger region of the invariant manifold is unraveled further apart from
the point~$\theta_{\rm unstab}(n)$.
When the integration time is sufficiently large,
the whole separatrix that joints two \emph{neighboring} hyperbolic points
is equally resolved
as the stable manifold 
emanating from the hyperbolic point~$\theta_{\rm unstab}(n)$
that is better identified with a forward-time evolution
coincides with the unstable manifold that emerges from
the hyperbolic point~$\theta_{\rm unstab}(n+1)$.

\subsubsection{Lorenz '63 System}
In this section, 
we present the results for the Lorenz'63 system
defined by Eqs.~\eqref{eq:lorenz_x}--\eqref{eq:lorenz_z}
for the classical parameters~$\sigma = 10$, $\rho = 28$, and $\beta = 8/3$~\cite{Lorenz1963},
which allow chaotic motion in this
higher dimensional system
thanks to the presence of a strange attractor.
Our analysis is performed using a $1600 \times 1200$ grid of 
ICs
spanning the domain $x \in [-20, 20]$, $y \in [-30, 30]$
on the $z \equiv z_0 = 27.0$ plane.


\begin{figure}[ht!]
    \centering
    \includegraphics[width=0.9\textwidth]{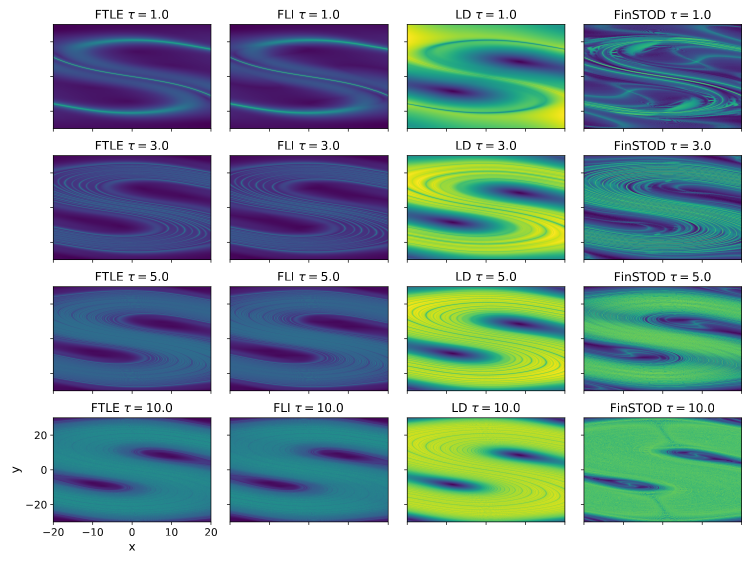}
    \caption{Same as Fig.~\ref{fig:pendulum_forw_raw}
    for the forward evolution of the Lorenz '63 system
    \eqref{eq:lorenz_x}-\eqref{eq:lorenz_z} on the $z=27.0$ 
    plane of initial conditions.
    }
    \label{fig:lorenz_4x4}
\end{figure}

Figure~\ref{fig:lorenz_4x4} displays the projection of the strange attractor 
onto the $z=27$ plane for increasing values of~$\tau$.
Already at short integration times ($\tau = 1$), all indicators reveal
the characteristic S-shaped skeleton that marks the boundary between
trajectories spiraling toward different lobes of the attractor.
As~$\tau$ increases, this structure becomes progressively more saturated,
reflecting the accumulation of stretching and folding
as the chaotic dynamics have more time to separate initially nearby trajectories.
FTLE, FLI, LD, and {\FinSTOD} show excellent structural agreement throughout the evolution,
with {\FinSTOD} capturing the same intricate filamentary patterns
as the other indicators.
This confirms that in this high-dimensional chaotic regime,
the ontological identity of trajectories is shaped by the same stretching
mechanisms that govern geometric separation.
The complete temporal evolution can be observed in the
\texttt{C\_lorenz\_*} video series of the repository~\cite{FinSTOD_repo},
where the progressive saturation of the attractor's structure is shown
frame by frame for all indicators.

\subsection{Non-autonomous systems}
In this section,
we present the {\FinSTOD} results for the non-autonomous systems
described in Sec.~\ref{sec:systems_non-autonomous}.

\subsubsection{Forced Duffing Oscillator}
The forced Duffing oscillator, governed
by~\eqref{eq:duffing_x} and~\eqref{eq:duffing_v}, introduces time-dependency and a complex fractal basin structure. 
Here,
we 
study
both forward and backward time evolutions for the set of parameters
$\alpha = -1.0$, $\beta = 1.0$, $\delta = 0.3$, $\gamma = 0.5$, and $\Omega = 1.2$,
where the \emph{double-well} dynamics are observed.
The analysis is performed on a phase-space grid of
$1000 \times 1000$ ICs
defined over the domain $x \in [-2, 2]$, $v \in [-2, 2]$.
The system is examined 
over a total time interval of $[0, 20.0]$.

\begin{figure}[ht!]
    \centering
    \includegraphics[width=0.8\textwidth]{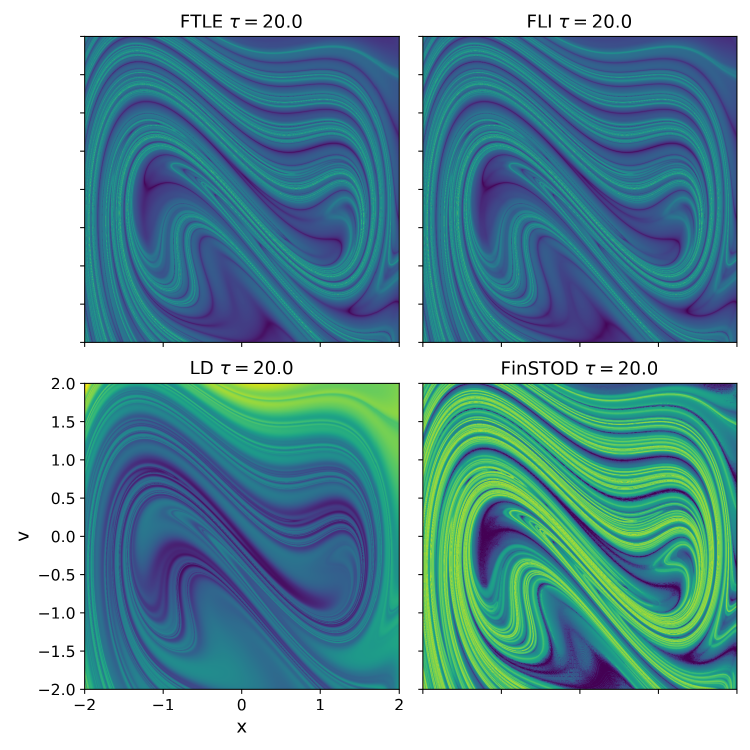}
    \caption{Same as Fig.~\ref{fig:pendulum_forw_raw}
     for the Duffing oscillator
     \eqref{eq:duffing_x}-\eqref{eq:duffing_v}
      at an integration time $\tau=20$ (forward time).
}
    \label{fig:duffing_forw_2x2}
\end{figure}

We begin with the forward-time analysis.
Figure~\ref{fig:duffing_forw_2x2} 
compares the results of the different indicators considered up to an integration time of
$\tau=20$,
corresponding to a time window that contains five complete periods of the external forcing..
Remarkably,
there is an excellent agreement 
between the {\FinSTOD} results and those related the studied standard indicators
(FTLE, FLI, and LD).
In all cases,
the repelling
LCS that play the role of the unstable manifolds 
are clearly identified.
They act as true transport barriers;
an IC starting in their close vicinity is pushed away
toward one of the two potential wells.
Moreover,
{\FinSTOD} not only resolves the same intricate fractal boundaries as the FTLE, FLI, and LD but does so with even greater detail and contrast, resolving fine filaments that appear as soft gradients in the standard methods.

\begin{figure}[ht!]
    \centering
    \includegraphics[width=0.9\textwidth]{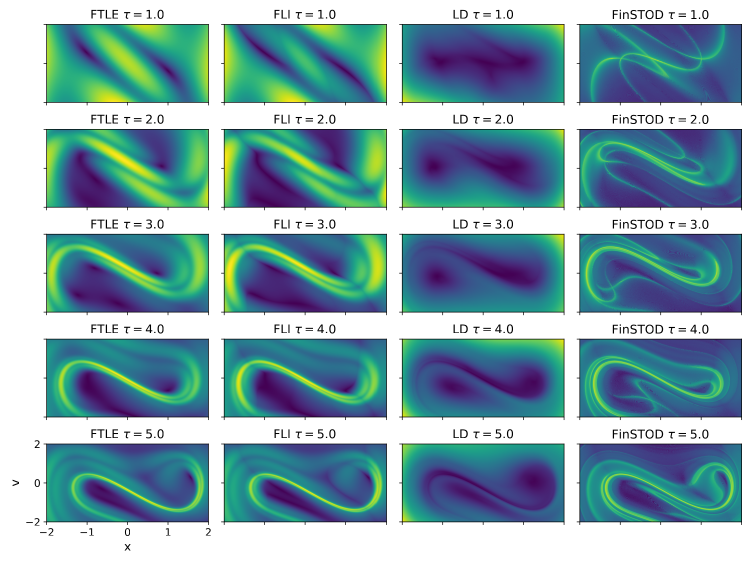}
    \caption{Same as Fig.~\ref{fig:duffing_forw_2x2} for $\tau \in [1, 5]$.}
    \label{fig:duffing_forw_4x4}
\end{figure}

\begin{figure}[ht!]
    \centering
    \includegraphics[width=0.9\textwidth]{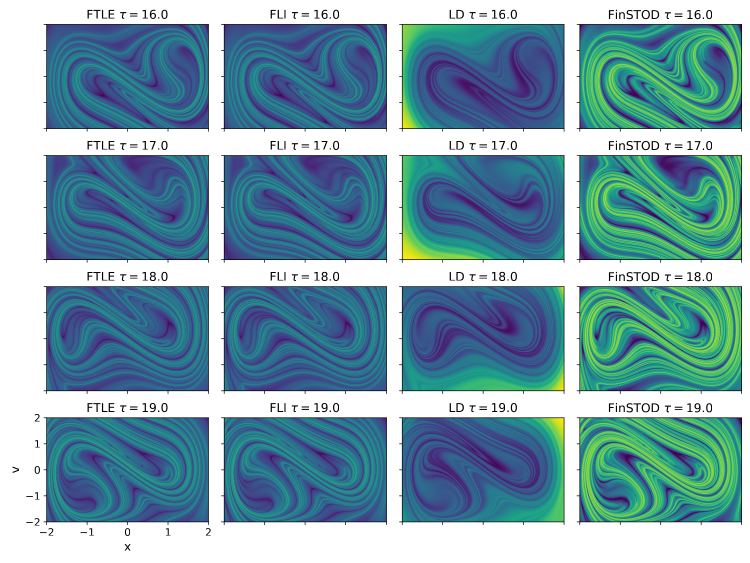}
    \caption{Same as Fig.~\ref{fig:duffing_forw_2x2} for $\tau \in [16, 19]$.}
    \label{fig:duffing_forw_5x4}
\end{figure}

As in the autonomous systems reported in Sec.~\ref{sec:systems_autonomous},
the repelling LCS
need an integration time 
sufficiently large to be resolved.
Thus,
in order to show this requirement,
we present in Figs.~\ref{fig:duffing_forw_4x4}
and~\ref{fig:duffing_forw_5x4}
the results exerted by the indicators considered in Fig.~\ref{fig:duffing_forw_2x2}
but for smaller values of~$\tau$.
As expected,
too low values of~$\tau$,
like the ones considered in Fig.~\ref{fig:duffing_forw_4x4},
 do not unveil any structure.
As a consequence,
the identification of the LCS that are responsible for the 
strongly chaotic dynamics requires longer integration times.
Indeed,
the results of Fig.~\ref{fig:duffing_forw_5x4},
which range from~$\tau = $16 to~19,
seem to have a suitable time, where the repelling LCS
become visible.
As can be inferred from the figure,
the shape of the LCS strongly depends on the~$\tau$ value
as they are time-dependent.
Moreover,
there is a consistent high-fidelity matching 
between the {\FinSTOD} results and those rendered by the reference indicators,
a result that shows the ability of {\FinSTOD} to also address non-autonomous systems.


\begin{figure}[ht!]
    \centering
    \includegraphics[width=0.9\textwidth]{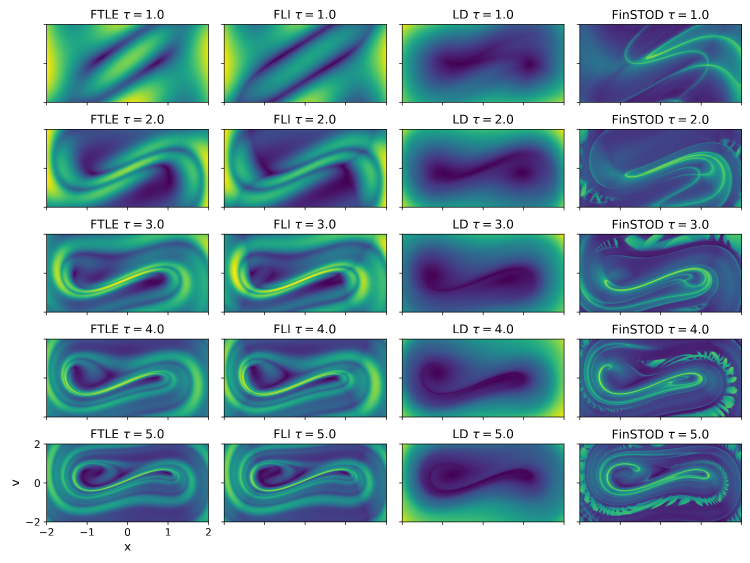}
    \caption{Same as Fig.~\ref{fig:duffing_forw_4x4} for the backward time evolution.}
    \label{fig:duffing_back_5x4}
\end{figure}

\begin{figure}[ht!]
    \centering
    \includegraphics[width=0.9\textwidth]{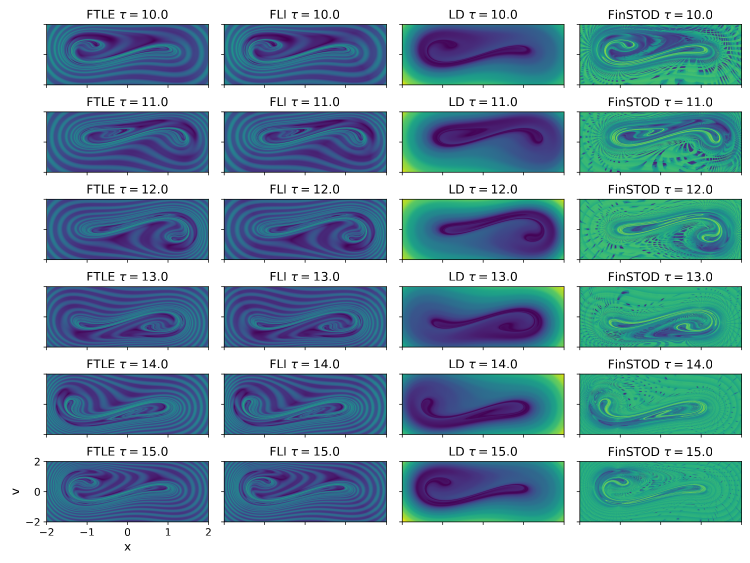}
    \caption{Same as Fig.~\ref{fig:duffing_forw_4x4} for the backward time evolution with $\tau \in [10, 15]$.}
    \label{fig:duffing_back_6x4}
\end{figure}

The backward-time analysis
(Figs.~\ref{fig:duffing_back_5x4}-\ref{fig:duffing_back_6x4}) yields analogous observations for the
attracting LCS.
Even for short-time integrations
(e.g., Fig.~\ref{fig:duffing_back_5x4}, where~$\tau \in [1, 5]$),
{\FinSTOD} begins to unveil these structures,
which function as the stable manifolds introduced in Sec.~\ref{sec:systems_autonomous}.
In forward time, trajectories initiated near
these LCS tend to converge and adhere to them,
confirming their role as the 
\emph{skeleton} of the attractor.


As the system evolves for longer integration times,
the {\FinSTOD} is still able to track the attracting the LCS with high definition
but not with the superb performance observed in the forward-time evolutions
previously discussed.
This is clearly seen for the medium integration times considered in Fig.~\ref{fig:duffing_back_6x4},
which range from~$\tau = 10$ to~15.
In this instance,
the LCS surrounding the potential minima are correctly identified.
However,
at larger separations from these points,
the {\FinSTOD} seems to saturate, and the LCS do not present such a clearly recognizable pattern
as in the FTLE and FLI cases,
but the performance seems to be better than in the case of the LD-plots.

\begin{figure}[ht!]
    \centering
    \includegraphics[width=0.8\textwidth]{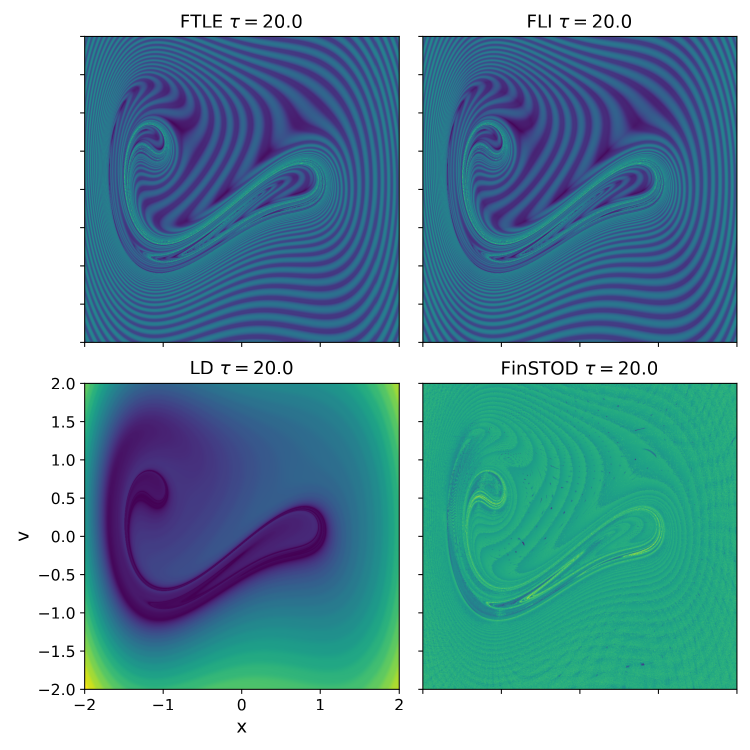}
    \caption{Same as Fig.~\ref{fig:duffing_forw_2x2} for the backward time evolution.}
    \label{fig:duffing_back_tau20}
\end{figure}

To conclude,
we present in Fig.~\ref{fig:duffing_back_tau20}
the results for~$\tau=20$.
In this case, the {\FinSTOD} is more strongly saturated.
Still, {\FinSTOD} remains highly effective at identifying the primary ridges and ripples resolved by the standard indicators.
We attribute the observed saturation to the increasing complexity of path identities at longer integration times, which leads to a more saturated scalar field in the {\FinSTOD} results.
The compared evolution for both orientations can be seen in the repository videos
\texttt{D\_duffing\_*}~\cite{FinSTOD_repo}.

\subsubsection{Time-Dependent Double Gyre}

In this section,
we discuss the {\FinSTOD} results for
forward-in-time evolution
of the double gyre~\eqref{eq:dg_u}--\eqref{eq:dg_v}
with the parameters:
$A = 0.1$, $\epsilon = 0.25$, and
$\kappa = \pi/5$. 
All computations have been conducted
on a $2000 \times 1000$ grid of 
ICs over the domain $x \in [0, 2]$, $y \in [0, 1]$
and
for~$\tau \in [0, 15]$.
Due to the inherent symmetry of the system's velocity field~\eqref{eq:dg_u}-\eqref{eq:dg_v},
we omit the results for the backward time evolution as they show
a high degree of structural similarity to the forward-in-time results presented. 
Specifically,
the backward evolution effectively corresponds to a vertically flipped version of the forward case
\cite{Shadden2005}.
%

\begin{figure}[ht!]
    \centering
    \includegraphics[width=0.9\textwidth]{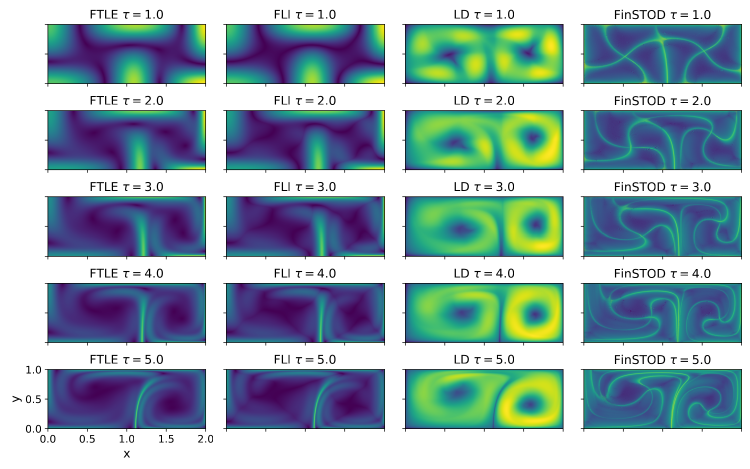}
    \caption{Forward evolution of the double gyre~\eqref{eq:dg_u}--\eqref{eq:dg_v}
                      for $\tau \in [1, 5]$.}
    \label{fig:dg_forw_grid_A}
\end{figure}
\begin{figure}[ht!]
    \centering
    \includegraphics[width=0.9\textwidth]{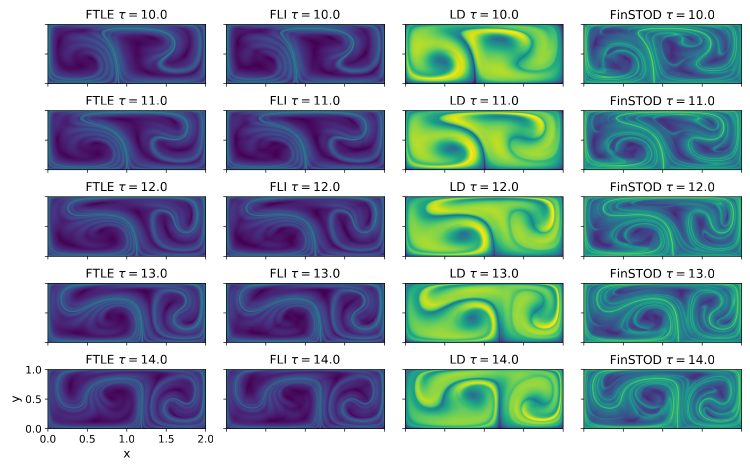}
    \caption{Same as Fig.~\ref{fig:dg_forw_grid_A} for $\tau \in [10, 14]$.}
    \label{fig:dg_forw_grid_B}
\end{figure}

Figure~\ref{fig:dg_forw_grid_A} shows the results for $\tau \in [11, 15]$.
For the shortest integration time ($\tau = 1$),
the FTLE- and FLI-plots are organized in a set of \emph{cells}
that are separated by LCS.
The LCS can be slightly noticed in the LD, but in a much smaller degree.
On the contrary,
the LCS that determines the dynamics are clearly visible on the {\FinSTOD}-plot.
In other words, 
the {\FinSTOD} consistently identifies the nascent skeleton of the flow.
As can be inferred by comparison with the remaining integration times,
the structure observed changes with the integration time.
Noticeably,
in all examples considered the {\FinSTOD} have a much more complex structure
that the remaining indicators,
which, for the integration times considered (especially for $\tau = 5$) 
only unravel the LCS that emerges almost vertically from 
the $x$ axis at~$x \approx 1.1$.
Thus,
what the standard indicators only timidly hint at as shadowy or coarse internal structures, {\FinSTOD} fully resolves.
The structure of the LCS becomes more complex as the value of~$\tau$ 
is further increased.
Note that the LCS gets well resolved 
over a larger range in Fig.~\ref{fig:dg_forw_grid_B}.
The results of FTLE, FLI and LD are consistent with those inferred with {\FinSTOD}.
Remarkably, 
FTLE and FLI show many more ripples than the LD, which presents a blurry behavior.
Conversely,
 {\FinSTOD} provides a high-definition plot.
Noticeably,
these ripples are better identified with {\FinSTOD},
which also unravel other filamentary structures that remain hidden or unresolved 
for the standard indicators.
This higher sensitivity indicates that {\FinSTOD} provides a more exhaustive probe into the system's instability, identifying regions of high divergence that traditional indicators may overlook.

As with the previous chaotic systems, the compared evolution can be seen in the \texttt{E\_doublegyre\_*} repository videos~\cite{FinSTOD_repo}, and the single raw {\FinSTOD} evolution shows the main structure of the double gyre singled out in contrast.


\section{Conclusions}
\label{sec:conclusions}

In this work, we have introduced Strong Trajectorial Ontological Differentiation (STOD), a novel trajectory-based framework that reframes dynamical divergence through the lens of ontological similarity~\cite{GarciaCuadrillero2026}.
By treating the complete trajectory of a particle as the unique identity of its initial state, STOD quantifies the similarity between trajectories using a component-wise cancellation process. While the STOD framework allows for various implementations, this study has focused on the {\FinSTOD} variant, which analyzes orbits in reverse order from their first states. Our local application of {\FinSTOD} provides a visualization of the flow's internal boundaries that complements and, in many regimes, offers distinct advantages compared to standard indicators like the
FTLE and the LD.

A primary conclusion of this study is that the STOD framework, and specifically the {\FinSTOD} variant, is robust across varying levels of system complexity. In simple linear regimes, where variational indicators like FTLE and FLI may be less informative due to spatially uniform stretching, {\FinSTOD} successfully resolves the underlying geometry. This aligns {\FinSTOD} with the primary strengths of Lagrangian Descriptors, which have traditionally been the preferred tool for such systems. However, {\FinSTOD} offers a notable alternative by complementing the strengths of both LD and FTLE: it maintains the robustness of the former while providing the high-resolution, filamentary detail typically associated with the latter. Unlike the relatively coarse ridges often produced by LD, {\FinSTOD} generates sharp, emphatic lines that resolve the fine structure of the chaotic sea with high precision.

Furthermore, {\FinSTOD} proves to be fundamentally system-agnostic regarding the analytical methods of the flow. While FTLE requires the computation of Jacobians and LD relies on the existence of specific velocity integrals---operations that can be numerically compromised or physically uninformative depending on the system's nature---{\FinSTOD} only requires the existence of the system's dynamics itself,
although the requirement to store and compare complete trajectories
introduces a computational overhead compared to standard indicators. By focusing on the visitation patterns 
of trajectories on a reference grid rather than on local linearizations, {\FinSTOD} provides a direct and exhaustive probe into the system's stability. Indeed, our results in the pendulum and the time-dependent double gyre demonstrate that this trajectory-based logic identifies not only the primary structures but also nascent path connections and hidden filaments of high divergence that traditional indicators may overlook.

It is important to emphasize that {\FinSTOD} is just one of many possible applications of the STOD framework. Other variants, exploring different trajectory orientations and comparison rules, are already in progress, and we hope to present these results in the near future. In conclusion, STOD provides a fresh and mathematically distinct perspective on the analysis of dynamical systems. By shifting the focus from infinitesimal geometric separation to the ontological similarity of historical trajectories, we have developed a measure that is both a robust alternative to existing methods and a sensitive probe into the structures that govern complex flows. {\FinSTOD}, in particular, stands as an effective indicator, capable of revealing the underlying order of chaotic systems with clear resolution.

\section*{Acknowledgments}

Funding was provided by the PRIORITY grant (PID2021-127202NB-C22) to JAC, and 
by the grants PID2021-122711NB-C21
and PID2024-157869NB-I00
funded by MCIN / AEI / 10.13039 / 501100011033
and ``ERDF A way of making Europe''. 

The authors acknowledge computing resources at the Magerit Supercomputer of the
Universidad Polit\'{e}cnica de Madrid.

\section*{Declaration of generative AI and AI-assisted technologies in the writing process}
Generative AI tools were used to improve the clarity and readability of the manuscript.
All content generated using these tools was critically reviewed, verified, and edited by the authors,
who accept full responsibility for the final version of the article.


\bibliography{myreferences, myreferences_new, lb_licn}

\end{document}